# Role of interface reaction on resistive switching of Metal/*a*-TiO$_2$/Al RRAM devices

Hu Young Jeong and Jeong Yong Lee[a]

*Department of Materials Science and Engineering, KAIST, Daejeon 305-701, Korea*

Sung-Yool Choi[b]

*Electronics and Telecommunications Research Institute (ETRI)*

*161 Gajeong-Dong, Yuseong-Gu, Daejeon, 305-700, Korea*

For the clear understanding of the role of interface reaction between top metal electrode and titanium oxide layer, we investigated the effects of various top metals on the resistive switching in Metal/*a*-TiO$_2$/Al devices. The top Al device with the highest oxygen affinity showed the best memory performance, which is attributed to the fast formation of interfacial layer (Al-Ti-O), as confirmed by high resolution transmission electron microscopy and electron dispersive spectroscopy. Hence, we concluded that the interface layer, created by the redox reaction between top metal electrode and TiO$_2$ layer, plays a crucial role in bipolar resistive switching behaviors of metal/TiO$_2$/Al systems.

[a] Electronic mail: j.y.lee@kaist.ac.kr
[b] Electronic mail: sychoi@etri.re.kr

Recently, binary transition metal oxides such as NiO,[1] TiO$_2$,[2] and ZnO,[3] have attracted a great deal of interest as high potential next-generation nonvolatile memory (NVM) due to their simple device structure, easy fabrication process and high CMOS compatibility. For the advanced development of actual RRAM devices, the clear understanding of the basic mechanism is essential, but the driving mechanism of the resistance switching behaviors remains controversial. To clarify the underlying mechanism, many researchers have studied intensively the microscopic origins of the resistive switching in binary oxide films.[4,5] In particular, the effects of various metal electrodes on the resistive switching of various binary oxide thin films such as NiO[6] and ZrO$_2$[7] have been investigated to identify the role of interface. However, there exist few works regarding the role of the interface between an amorphous TiO$_2$ film and the top metal electrodes.

In this letter, we report the influence of top metal electrodes on the resistive switching properties in Metal/TiO$_2$/Al devices with the several kinds of metallic electrode such as Al, Cr, Mo, W, and Pt in order to support our resistive driving model. In the case of Al top electrode, which has the lowest oxide formation energy, the largest hysteresis and highest on/off ratio were shown among the tested device structures. It can be strongly confirmed that the resistive switching behavior is deeply related with the interface reaction between top metal electrode and titanium oxide layer.

Amorphous TiO$_2$ films were deposited on Al/SiO$_2$/Si substrates by plasma-enhanced atomic layer deposition (PEALD) methods.[8,9] During the deposition process (200cycle), substrate temperature was kept at 180℃. Titanium tetra-iso-propoxide (TTIP) and oxygen plasma were used as the Ti and oxygen precursors, respectively. The aluminum electrodes with a thickness of 60nm were deposited by thermal evaporation method, forming the cross-bar type structures using a metal shadow mask with a line width of 60 μm, as shown in the right inset of Fig. 1(a). In order to investigate the electrical properties of amorphous titanium oxide films with several different metals, metal top electrodes (Al, Cr, Mo, W, and Pt) were deposited by thermal evaporation, e-beam evaporation, and RF magnetron sputtering methods. The electrical property (I-V curve) was measured using a Keithley 4200 Semiconductor Characterization System in a dc sweep mode. The cross-sectional HR-TEM images of Metal/TiO$_2$/Al samples were examined using a 300 kV JEOL JEM 3010 with a 0.17 nm point resolution. Annular dark field (ADF) scanning TEM (STEM) image and point energy dispersive x-ray spectroscopy (EDS) data were obtained using field emission transmission electron microscope (Tecnai G2 F30) equipped with STEM mode and an EDS analyzer.

Figure 1 shows the typical J-V curves of Metal/TiO$_2$/Al memory cells with Al, Cr, Mo, W, and Pt electrodes, which were measured in a voltage region (-4V~ 4V) under the dc voltage sweep. A Pt/TiO$_2$/Al device showed an ohmic behavior at the very low-voltage region (-0.1V ~ +0.1V) (data not shown here). This interesting phenomenon cannot be explained by a general semiconductor concept because Pt is expected to form a Schottky-like contact with the n-type semiconductor TiO$_2$ due to a high work function of Pt (~5.65eV).[10] On the other hand, a device with Al top metal of a low work function (~4.28eV)[10] represents the most non-ohmic behaviors, showing remarkably low current level in the negative voltage region. Therefore, it can be inferred that the new factor such as interface layer, which has a high resistance than TiO$_2$ itself, is involved in this system rather than the general theory of metal-semiconductor contact. All devices showed a hysteretic behavior and bipolar resistive memory switching. However, it is noted that the switching voltage, the current level and On/Off ratio varied. The Al/TiO$_2$/Al device shows considerably the highest hysteresis and On/Off ratio than any other top electrode devices, supporting that the resistance change is deeply associated with the top interface reactions between top electrodes and TiO$_2$ films.[8,9]

To understand the exact conduction mechanisms, the current-voltage relationships of I-V curves such as a double log plot were investigated. Figure 2(b) is the double-logarithmic plots of the J-V curves for the negative voltage regions in the case of various top electrodes (Al, Mo, and Pt). All devices exhibited similar I-V characteristics, which are well described by the trap-controlled space-charge limited current theory, as explained in our previous result.[8] However, three remarkable differences with the change of top metal electrodes are shown in the black dot-circle regions of Fig. 2(b). Firstly, in the region 2, metal electrodes having large oxygen affinity like Al and Mo showed the large exponents (n ~ 4) of V during the first negative sweep, whereas the Pt top electrode device exhibited the general square law dependence (n ~ 2) on voltage. Secondly, the maximum slope of log J-log V in the region 3, which was involved with ON/OFF ratio, increased as the oxide formation energy decreased. Another important difference was also observed in the slope of region 4 of the LRS. Al and Mo devices had lower values (~ 1.5) than Pt device (~ 2). The reason for the differences might be attributed to the interface layer formed at the vicinity of the top metal electrode during metal deposition process, as presented by our earlier works.[8,9] We suggested that the reversible formation and dissociation processes of the Al-Ti-O interface by the migration of oxygen ions under an applied bias played an important role in Al/$a$-TiO$_2$/Al devices. The higher slope of region 2 and 3 in the case of Al indicates the formation of more insulating interface layer (Al-Ti-O), which contains a large amount of oxygen ions, via the movement of oxygen ions present in bulk titanium oxide, thus resulting in the steep profile of oxygen ions. The accumulated oxygen ions move back to inner titanium oxide layer under the high electric field, making the device the On state. On the other hand, the low value of Pt electrode device means no generation of the insulating interface layer such as Pt-Ti-O. Negatively charged oxygen ions diffuse out to Pt electrode due to the absence of blocking layer, thus making a slow gradient in the distribution of oxygen ions at the vicinity of top interface region.

To directly demonstrate the abovementioned interface regions formed between various top metal electrodes and titanium oxide thin films, we performed the HRTEM measurements of each sample. Figure 2 is the cross-sectional HRTEM images of Al (Mo, Pt)/TiO$_2$/Al stacked devices taken with the same magnification. Figure 2(a) shows the image of the TiO$_2$/Al structure before the deposition of top metal electrode. Compared to Fig. 2(a), after the deposition of the Al top metal, a 3-nm-thick amorphous interface layer emerged, indicating the redox reaction between the Al and TiO$_2$ film. Compared with the Al device, the samples with Mo and Pt top electrodes did not show noticeable interface regions, indicating no dynamic reaction. However, it is remarkable that there was a change of contrast in titanium oxide layer itself. As the oxygen affinity of top metals decreased, a lighter contrast was observed, meaning the compositional change.

To further characterize this phenomenon, we performed a scanning TEM (STEM) point energy dispersive x-ray spectroscopy (EDS) measurement. In order to effectively investigate the effect of top metals, the specimens with a-TiO$_2$ thin films of 15 nm thickness deposited on Si substrates by a PEALD method were prepared, following different metal depositions (Al and Pt). For the detailed investigation of a chemical composition of each titanium oxide thin layer, a point EDS analysis was conducted using a STEM with a 0.7 nm probe size. Figure 4 shows annular dark field (ADF) STEM images of three different samples (*a*-TiO$_2$/Si, Al/*a*-TiO$_2$/Si, and Pt/*a*-TiO$_2$/Si), point EDS spectra measured at the region marked with red circle, and EDS line profiles scanned from a Si substrate (point S) to a surface region (point F). The most striking feature was the change of chemical composition of titanium oxide thin films. The atomic ratio of Ti and O, seen in the right inset of Fig. 3a, b, and c, indicates that a large amount of oxygen vacancies were created in the case of the Pt top electrode, which was well consistent with the EDS line scan profile, as shown in Fig. 3d, e, and f. This interesting result is not agreement with the general belief that the reactive metal such as Al can easily gather oxygen ions on the titanium oxide layer due to highly negative oxide formation energy. Thus, in our system, another model has to be considered. Schmiedl et al.[11] demonstrated that the grain boundary diffusion of oxygen through a Pt layer of up to 17 nm is possible when the layer was exposed to an air at room temperature. In a real NVM device (Pt/Ba$_{0.7}$Sr$_{0.3}$TiO$_3$/SRO), the degradation of an endurance performance due to the diffusion of oxygen ions along the Pt grain boundaries was suggested when the negative bias was applied on the Pt top electrode.[12]

Based on these reports and our TEM results, the new schematic model to explain the compositional change of titanium oxide layer according to each top electrode was schematically sketched in Figure 4. In the case of the Al top electrode, the insulating layer (Al-Ti-O) is rapidly formed at the initial stage of metal growth owing to a strong oxygen withdrawing power of the Al metal. Because of this insulating barrier, there is no additional out-diffusion of oxygen ions within the titanium oxide layer, resulting in Al-Ti-O domain and corresponding $a$-TiO$_{2-X}$ domain, respectively. In contrast, the Pt metal cannot create the blocking layer such as Pt-Ti-O, even though it has a little power gathering oxygen ions from the titanium oxide layer. However, the oxygen ions continuously diffuse out to a surface through a grain boundary until the thickness of Pt reaches at a critical point, making the $a$-TiO$_2$ film more oxygen-deficient $a$-TiO$_{2-X}$. The light contrast, shown in Figure 2d, indicates this phenomenon. As expected, the Mo top electrode, whose oxygen affinity is a value between the Al and Pt, exhibited an intermediate contrast, as shown in Figure 2c.

In summary, we have investigated the influence of top metal electrodes on the resistive switching properties in Metal/TiO$_2$/Al devices. Although the all devices showed

BRS behaviors, the higher on/off resistance ratio could be obtained in the Al top metal case with the lowest oxide formation energy. The highest oxygen affinity induced the strong interaction between top metal electrode and the top domain of $TiO_2$ at the initial stage of metal deposition process, thus creating newly formed interface layer, as confirmed by HRTEM images and STEM EDS analysis. On the base of these results, it is strongly demonstrated the switching behaviors of metal/$TiO_2$/Al memory devices depend on the interface reaction with top metal electrodes rather than a work function.

**Acknowledgments**

This work was supported by the Next-generation Non-volatile Memory Program of the Ministry of Knowledge Economy (10029953-2009-31) and the Korea Research Foundation Grant (MOEHRD) (KRF-2008-005-J00902). The authors sincerely thank the technical staffs of the Process Development Team at ETRI for their support with the PEALD facility.

**Figure Captions**

FIG. 1. (Color online) (a) Typical J-V curves of Al (Cr, Mo, W, and Pt)/TiO$_2$/Al. The right below inset shows a schematic diagram of our crossbar type memory with 60 μm ×60 μm (3600 μm$^2$). (b) J-V characteristics of Metal/TiO$_2$/Al devices in a double-logarithmic plot. The colored values mean the slopes of linear fitting of each separated regions.

FIG. 2. (Color online) Cross-sectional HRTEM images with the same magnification of (a) TiO$_2$/Al layered structure before the deposition of top metal electrodes, (b) Al/TiO$_2$/Al device, (c) Mo/TiO$_2$/Al, ad (d) Pt/TiO$_2$/Al. The red dot lines have all the same width in the each figure to compare interface changes after deposition of top metal electrodes.

FiG. 3. (Color online) Cross-sectional annular dark field (ADF) STEM Z-contrast image of (a) *a*-TiO$_2$/Si, (b) Al/*a*-TiO$_2$/Si, and (c) Pt/*a*-TiO$_2$/Si samples. The point EDS spectra are shown in the right inset of (a), (b) and (c). The line profile of each sample are displayed in (d), (e), and (f), respectively.

FiG. 4. (Color online) Schematic illustration showing the different phenomena occurred at the top interface region during Al ((a) and (b)) and Pt ((c) and (d)) top metal deposition.

Figure 1. (Jeon et al.)

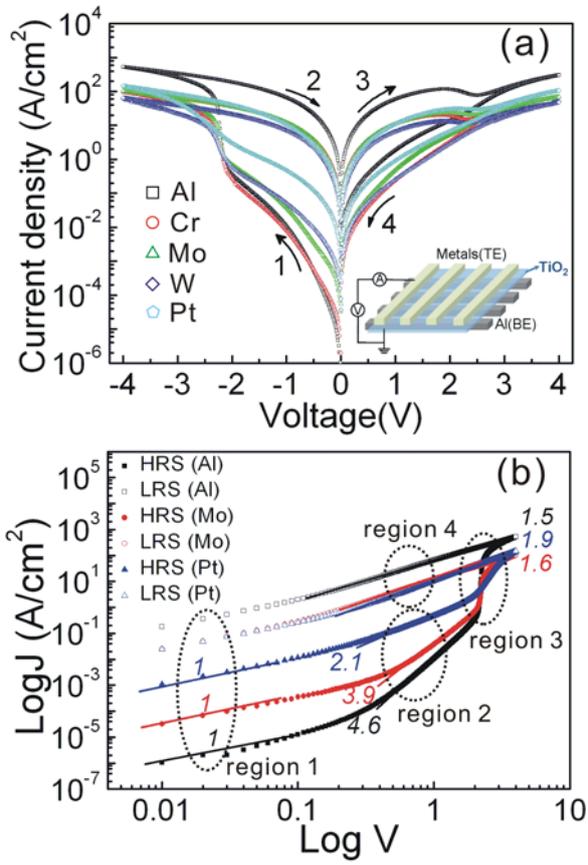

Figure 2. (Jeon et al.)

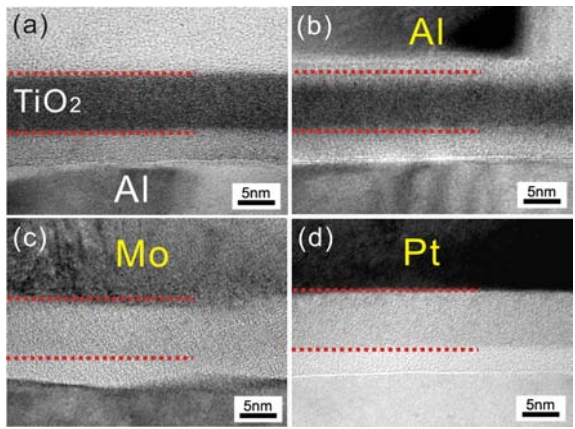

Figure 3. (Jeon et al.)

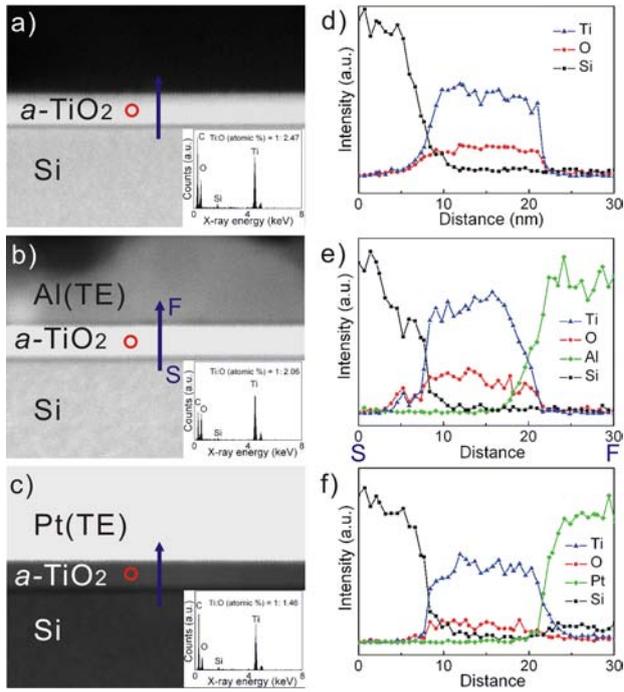

Figure 4. (Jeon et al.)

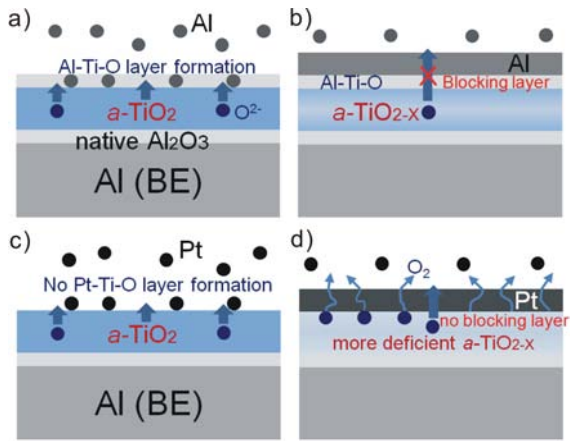